\documentclass{article}
\usepackage[T1]{fontenc}
\usepackage[utf8]{inputenc}
\usepackage{ismir} 
\usepackage{amsmath,cite,url}
\usepackage{graphicx}
\usepackage{color}

\usepackage{booktabs}
\usepackage{siunitx}
\usepackage{soul}
\usepackage{amssymb}

\def\ourDataset{\emph{AllMusicCaps}}
\def\ourShort{\emph{AMC}}
\def\ourQuotes{\emph{AMCQuotes}}
\def\ourStruct{\emph{AMCStruct}}

\title{AllMusicCaps: Album Reviews as Complementary Supervision for Music CLAP}
 
\multauthor
  {Pablo Alonso-Jim{\'e}nez \hspace{0.5cm} Xavier Lizarraga-Seijas \hspace{0.5cm} Xavier Serra \hspace{0.5cm} Dmitry Bogdanov}
  {Music Technology Group, Universitat Pompeu Fabra, Barcelona \\
  {\tt\small pablo.alonso.ji@gmail.com}
  }

\def\authorname{P. Alonso-Jim{\'e}nez, X. Lizarraga-Seijas, X. Serra, and D. Bogdanov}

\begin{document}

\maketitle

\begin{abstract}
Recent open text-audio contrastive models (CLAPs) are typically trained with LLM-generated captions derived from tag datasets or web search results, which tend to be accurate but expressively narrow.
As a complementary source, we explore human-written album reviews, specifically expert reviews from AllMusic: they exist at scale and 
carry narrative cues, evaluative adjectives, and scene framing that other sources lack.
Since raw reviews are too noisy for direct use as captions, we first build a caption corpus with \num{245346} samples via an LLM preprocessing pipeline that identifies descriptive musical quotes and rewrites them into training-ready captions.
We find that album review supervision yields the largest retrieval gains on a human-written caption benchmark (Song Describer), particularly for complex queries that other existing caption datasets leave uncovered.
In addition, we revisit the training recipe and show that SigReg regularization, which encourages an isotropic Gaussian distribution in the embedding space, improves MLP probing across classification tasks, as well as text-to-music retrieval.
The resulting model outperforms open CLAP-style baselines on text-to-music retrieval, zero-shot classification, and most MLP probing tasks.
We release the review-derived caption dataset and model weights to support future research.
\end{abstract}

\section{Introduction}\label{sec:introduction}

Joint text-audio representation models are central to a wide range of Music Information Retrieval tasks.
They enable multimodal retrieval~\cite{huang2022mulan}, conditioned generation~\cite{agostinelli2023musiclm}, semantic similarity evaluation~\cite{huang2023make}, and discriminative tasks through zero-shot or transfer learning~\cite{elizalde2023clap, ramoneda2026benchmarking}.
Recently, LLMs have been widely used to generate paired audio-caption training datasets.
Popular approaches include obtaining captions from existing tag datasets~\cite{huang2023noise2music, doh2023lpmusiccaps, roy2025jamendomaxcaps}, and using retrieval-augmented generation over web searches~\cite{wu2025clamp}.
However, the captions these pipelines produce, while usually accurate and objective, do not capture complex musical queries featuring narrative framing (e.g.\ ``feels like\ldots'') or evaluative adjectives (e.g.\ ``raspy vocals'').

To overcome this limitation, we propose employing album reviews available at AllMusic,\footnote{\url{https://www.allmusic.com/}} an online music database and review website already considered in MIR research~\cite{hu2012genre, schindler2012facilitating, schreiber2015improving, bogdanov2019acousticbrainz}.
AllMusic reviews are written by experts, exist at scale, and contain information that goes beyond tag-like information.
However, reviews are not directly usable as supervision: a single album review may discuss several tracks, drift into editorial commentary, or describe context unrelated to the audio.
To this end, we compare two LLM-based caption generation strategies, producing \ourDataset\ (\ourShort), a music caption dataset suitable for training text-music models 
covering a blind spot left by current caption datasets. 


Beyond the data, we investigate the effect of using different layers of the audio encoder and compare training objectives.
In particular, we experiment with replacing the standard InfoNCE with the Sigmoid loss~\cite{zhai2023sigmoid}, and with adapting LeJEPA~\cite{balestriero2025lejepa}, a self-supervised method that encourages isotropic Gaussian representations through the SigReg regularizer.
Our results show that reviews alone are insufficient to match the diversity of the existing music and sound corpora.
However, when combined with those corpora, they contribute as a complementary signal, with the largest gains concentrated on high-complexity captions that mix narrative, affect, and figurative description.


The main contributions of our work are the following:
\begin{enumerate}
    \item We introduce \ourDataset, a new music caption dataset derived from AllMusic expert album reviews enriched with YouTube and Discogs metadata, suitable for multimodal (text-audio) research in music.

    \item We study the audio-encoder layer selection and training objective, and consolidate the findings into a recipe that improves over the default setup across retrieval, zero-shot classification, and probing.

    \item We find that SigReg regularization benefits text-to-music retrieval and probing under certain conditions.

\end{enumerate}

All materials derived from this work, including the \ourDataset\ dataset, model weights, and code, are released for non-commercial scientific research purposes only.
\footnote{\url{https://github.com/mtg/allmusiccaps/}}

\begin{figure*}[ht]
    \centering
    \includegraphics[width=\linewidth]{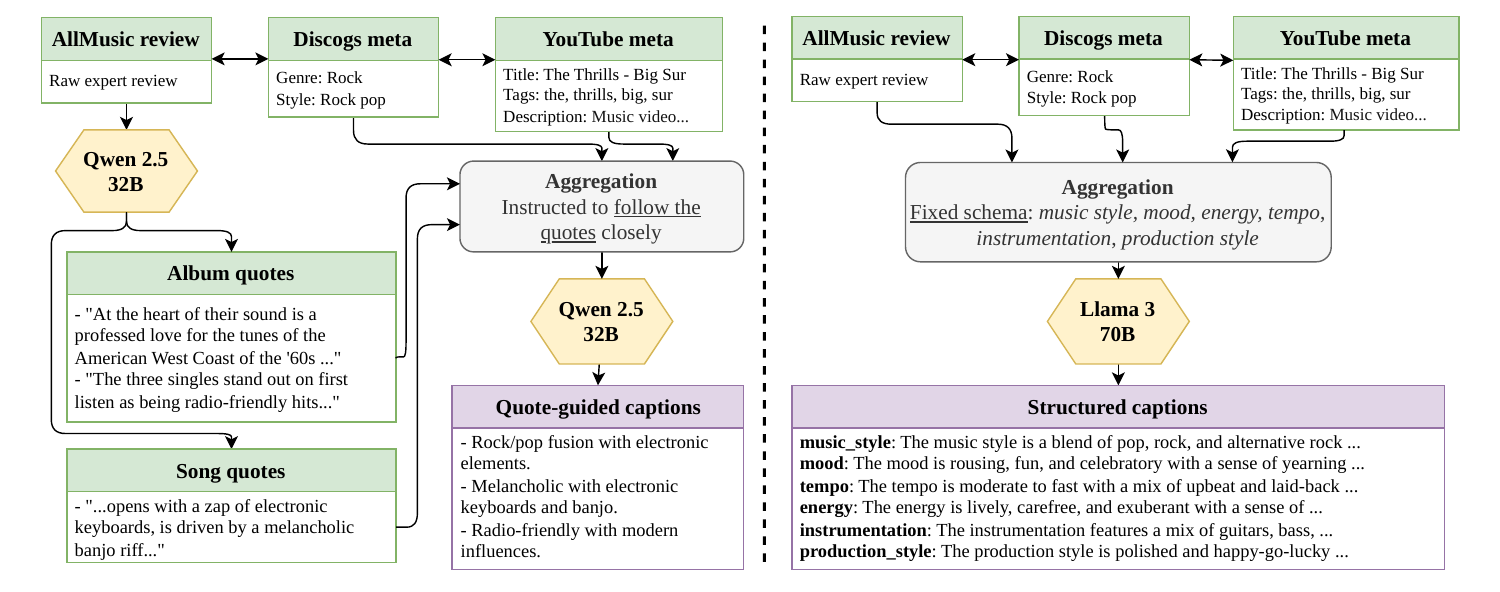}
    \caption{
    \ourDataset\ caption generation example for the song \textit{``Big Sur''} by \textit{The Thrills}.
    \textbf{Left:} Quote extraction followed by track-level information aggregation and rewriting. \textbf{Right:} Single-step structured caption generation.
    }
    \label{fig:diagram}
\end{figure*}

\section{Related work}\label{sec:related_work}

\textbf{Music text-audio supervision for contrastive models.}
Music CLAPs trace back to general-purpose text-audio contrastive models trained on tag datasets with mixed coverage of general-audio and music~\cite{favory2020coala, wu2023largescale, elizalde2023clap}.
    The first music-focused CLAPs draw text from professional captions on private catalogs (MusCALL~\cite{manco2022contrastive}), weakly-associated descriptions, tags, and playlists from YouTube (MuLan~\cite{huang2022mulan}), or music tagging datasets (TTMR~\cite{doh2023toward}).
CLaMP3~\cite{wu2025clamp} relies on web search over song title and artist name with LLM post-processing.
Across existing systems, album-review text has not been explicitly used at scale.


\textbf{Caption generation for music.} LP-MusicCaps~\cite{doh2023lpmusiccaps} prompted an LLM to convert tag sets into natural-language captions, providing the dominant source of text-music supervision.
Follow-ups refine prompting~\cite{doh2024enriching} or add retrieval-augmented generation from web search results~\cite{wu2025clamp}.
Our pipelines differ in the underlying source: a large-scale corpus of expert-written album reviews, rather than tags or open-web retrieval.


\textbf{Self-supervised representation learning.} Most joint-embedding approaches adapt self-supervised objectives to the multimodal case.
For example, CLAPs~\cite{elizalde2023clap,wu2023largescale} rely on the InfoNCE loss~\cite{oord2018representation}, while SLAP~\cite{guinot2025slap} adapts BYOL~\cite{grill2020bootstrap} for negative-free joint-embedding pretraining.
In addition, other methods combine contrastive with self-supervised objectives, such as masked token modeling~\cite{niizumi2024m2d}.
We propose combining LeJEPA's SigReg regularizer~\cite{balestriero2025lejepa} with the InfoNCE loss, which recovers the probing benefits of isotropic Gaussian embeddings without the retrieval performance cost incurred by negative-free training.

\section{Data}\label{sec:data}

Our starting point is a collection of \num{94041} AllMusic album reviews paired to Discogs release metadata by matching artist and album names.
We extract the YouTube links associated with each Discogs release from the Discogs monthly data dump\footnote{\url{https://data.discogs.com/}} and retrieve the corresponding audio clips and metadata.
The matching process yields \num{245346} audio tracks paired to metadata at the album level (AllMusic reviews and Discogs editorial information) and track level (YouTube metadata).
Since most of the review text does not directly describe the audio, we design and compare two pipelines that distill the relevant information and combine the different metadata sources into training captions.
Both pipelines are summarized in Figure~\ref{fig:diagram}.

\subsection{Quotes vs. structured captions}

\textbf{\ourQuotes.} The goal of this pipeline is to stay close to the reviewer's stated observations and minimize hallucinations, at the cost of coverage. It operates in two stages, both using Qwen~2.5-32B.

\emph{Stage~1, extraction.} The first LLM receives the full AllMusic review and is prompted to extract two types of verbatim spans: \texttt{album quotes}, a list of text segments describing the sound of the album as a whole, and \texttt{song quotes}, a dictionary mapping track titles to text segments describing specific songs.
The system prompt restricts the extractor to factual descriptions of musical and acoustic characteristics (genre, production style, instrumentation, mood, rhythm) and explicitly excludes subjective opinions, biographical information, release dates, commercial performance, and album or artist names.

\emph{Stage~2, caption generation.} Raw quotes are not yet usable as training captions: most of the extracted captions are album-level comments rather than track-specific (e.g., a ballad within a rock album may not fit the overall album description), and they carry the review's narrative framing.
The second LLM 
combines the information available for each track and rewrites it into self-contained captions.
For each track, we include the respective \texttt{album quotes}, the Discogs release style and genre tags, and the YouTube description and tags.
Additionally, we match the YouTube title against the review's \texttt{song quotes} track titles using a case-insensitive substring regex, and attach the corresponding quotes when there is a single positive match.
The LLM is asked to return a list of short sentences ($<\!15$ words) prioritizing YouTube tags and song quotes over album-level metadata when these conflict.

\textbf{\ourStruct.} The goal of the structured captions pipeline is the opposite: maximize coverage of the aspects of interest for retrieval.
A single LLM is prompted with all the metadata available (AllMusic album review, YouTube description and tags, and Discogs genre and style tags) and a fixed schema of musical attributes (music style, mood, energy, tempo, instrumentation, and production style) and is asked to fill in each field, inspired by M4-RAG's structured captioning strategy~\cite{wu2025clamp}.
Lacking a quote-extraction stage to narrow the input, this pipeline must digest the raw metadata in one pass, so we allocate it more capacity (Llama~3-70B).
The LLM is instructed to infer fields from context when possible and leave them empty otherwise.
This yields uniform, dense captions at the cost of a higher hallucination risk, which manual inspection suggested is low. Its quantification was left for future work.

Both pipelines run on the same pool of tracks.
Outputs are validated for non-empty content and parseable YAML, and invalid generations are dropped, yielding a common subset of \num{245346} tracks with both types of captions.
To promote reproducibility and further multimodal text-music research, we publish the audio-caption pairs in the \ourDataset\ dataset.

\subsection{Additional caption datasets}

In addition to training models on our review-derived captions, we experiment with combining them with existing open caption datasets.
We consider two music datasets. 
LP-MusicCaps (\emph{LPMC})~\cite{doh2023lpmusiccaps} contains pseudo-captions generated from the Million Song Dataset/Last.fm tags~\cite{bertin2011msd}.
We consider the \emph{pseudo\_caption} and \emph{tag\_list} fields.
We additionally use \emph{M4-RAG}, an LLM-generated structured-caption corpus
where captions are produced with retrieval-augmented generation over web searches for each song title.
We consider the fields \emph{description}, \emph{background}, \emph{analysis}, and \emph{scene}.
Neither release ships with audio, so we match their track identifiers with our in-house music collection, which recovers \num{476}\,k tracks for LPMC and \num{243}\,k for M4-RAG.
Since the considered music collections mostly cover professionally recorded music, we also consider two general sound collections containing single-instrument and performance recordings, among other sources.
We observed that adding this data benefited our models in preliminary experiments.
Freesound (\emph{FS})~\cite{font2013freesound} contributes \num{327}\,k sounds downloaded from the Freesound API, filtered by duration (between 1 and 30 seconds), from which we use the \emph{description}, \emph{tags}, and \emph{category} fields as text. Pro Sound Effects (\emph{PSE})\footnote{\url{https://www.prosoundeffects.com}} contributes \num{1.02}\,M sounds, using its two-level taxonomy labels as text.
The interplay between these corpora is studied in Section~\ref{sec:experiments}.

\section{Method}\label{sec:method}

We use a two-tower CLAP model trained in two stages: a contrastive pre-training pass on raw metadata, followed by a training pass on the caption data from Section~\ref{sec:data}.

\subsection{Architecture}\label{sec:architecture}

We follow the standard two-tower CLAP setup~\cite{wu2023largescale}. The audio encoder is OMAR-RQ small~\cite{alonso2025omarrq}, a self-supervised conformer with \num{78}M parameters trained on \num{330}\,k hours of music.
The text encoder is the all-MPNet-base-v2 sentence transformer~\cite{song2020mpnet}, a text-embedding model with \num{110}M parameters.
We considered larger encoders in preliminary experiments, but under a fixed GPU memory budget, smaller encoders paired with larger batch sizes (more contrastive negatives) performed better.
The audio representations over \num{10} seconds are averaged into a single embedding, and the audio and text embeddings are then projected to a shared \num{512}-dimensional space with linear heads.

\subsection{Pre-training with raw metadata}\label{sec:pretraining}

Previous work found large-scale contrastive training using raw web-mined text to be beneficial for music CLAP models~\cite{huang2022mulan}.
Following that, we initialize both towers from the pre-trained weights of Section~\ref{sec:architecture} and then run a contrastive pre-training stage aiming to pre-align the two towers and reduce the training effort required in the main training phase.
We rely on a pool of \num{6.5}\,M audio tracks paired to raw YouTube and Discogs metadata that is used directly, with no LLM-generated captions.
We use a fixed subset of fields, from YouTube (description, categories, tags, view count) and from Discogs (labels, genres, styles, country, release date).
Fields are serialized as YAML and fed to the text encoder.
For YouTube tracks mapped to multiple Discogs releases, we sample one release per epoch, and randomly drop either the YouTube or the Discogs block with probability $p = 0.3$.
Our models optimize the symmetric InfoNCE loss~\cite{radford2021learning},
\begin{equation}
  \mathcal{L}_{\text{InfoNCE}} = \tfrac{1}{2}\!\left(\mathcal{L}_{a\to t} + \mathcal{L}_{t\to a}\right),
\label{eq:infonce}
\end{equation}
\begin{equation}
  \mathcal{L}_{a\to t} = -\frac{1}{N}\sum_{i=1}^{N}
    \log\frac{e^{\mathbf{a}_i^\top \mathbf{t}_i / \tau}}{\sum_{j=1}^{N} e^{\mathbf{a}_i^\top \mathbf{t}_j / \tau}},
\end{equation}
where $\mathbf{a}_i, \mathbf{t}_i \in \mathbb{R}^d$ are the $\ell_2$-normalized audio and text embeddings for sample $i$ in a batch of size $N$, $\tau > 0$ is a temperature parameter, and $\mathcal{L}_{t\to a}$ is defined symmetrically.

The models are trained for \num{400}\,k steps on \num{8} Nvidia H100 \SI{64}{\giga\byte} GPUs with a batch size of \num{64} \SI{10}{\second} segments at \SI{24}{\kilo\hertz} (negatives are drawn per device, not gathered across GPUs), using AdamW with a peak learning rate of \num{1e-4}, cosine annealing, and \num{20}\,k warm-up steps.
We use the final pre-trained checkpoint as the initialization for the captioned training stage described next.

\subsection{Training}\label{sec:training}

All models are initialized from the pre-trained checkpoint (Section~\ref{sec:pretraining}) and trained for \num{150}\,k steps on \num{8} Nvidia H100 \SI{64}{\giga\byte} GPUs with an effective batch size of \num{3072} \SI{10}{\second} segments at \SI{24}{\kilo\hertz}. We use the AdamW optimizer with a peak learning rate of \num{5e-5}, cosine annealing, and \num{15}\,k warm-up steps. Training takes approximately \num{36} hours per model.
All hyperparameters are held constant across Section~\ref{sec:experiments} unless stated otherwise, and all tables report the final \num{150}\,k-step checkpoint.

\textbf{Data processing and balancing.} In the captioned training stage, the five corpora span more than an order of magnitude in size, so we draw examples with a weighted sampler at target ratios \num{0.40} (\ourShort), \num{0.15} (LPMC), \num{0.25} (M4-RAG), \num{0.14} (FS), \num{0.06} (PSE), allocating \SI{80}{\percent} of each batch to music and \SI{20}{\percent} to general sound.
On the text side, each dataset implements a custom augmentation strategy:
\ourShort\ picks one release and joins up to \num{3} random sentences from its review-derived captions; M4-RAG samples one of four long-form fields (description, background, analysis, scene); LPMC picks the pseudo-caption or the Last.fm tag list with probability \num{0.5}; Freesound picks the user description with probability \num{0.2} and the tags otherwise; PSE shuffles the two-level taxonomy labels with a short file-name description.
To disambiguate terms shared across domains (e.g.\ \textit{rock} as a genre or an environmental sound), each text sample is prepended with an \texttt{[audio\_type\_token]}: \texttt{MUSIC} for \ourShort, LPMC, and M4-RAG, the \emph{broad category} for FS, and the \emph{parent category} for PSE, dropped at random in training so it is not required at inference.

\begin{table}
\centering
\footnotesize
\setlength{\tabcolsep}{3pt}
\begin{tabular}{p{2.5cm}cccc}
\toprule
 & \multicolumn{2}{c}{\textbf{Retrieval}} & \multicolumn{2}{c}{\textbf{ZS Class.}} \\
\cmidrule(lr){2-3} \cmidrule(lr){4-5}
\textbf{Data} & MuCaps & SongD. & GTZAN & FMA-S \\
 & MRR$\uparrow$ & MRR$\uparrow$ & Acc.$\uparrow$ & Acc.$\uparrow$ \\
\midrule
\textit{baseline} & 7.2 & 15.1 & 86.4 & 55.0 \\
\ourQuotes & 5.6 & 14.7 & 81.7 & 48.0 \\
\ourStruct & 4.6 & 15.4 & 84.3 & 47.5 \\
\textit{baseline+}\ourQuotes & \textbf{7.3} & \textbf{18.8} & \textbf{87.1} & \textbf{55.5} \\
\textit{baseline+}\ourStruct & \textbf{7.3} & 17.9 & 85.6 & 55.0 \\
\bottomrule
\end{tabular}
\caption{Downstream performance of each review-derived text corpus, the baseline (LPMC + M4-RAG + FS + PSE), and their combinations. 
\textbf{Bold} = best per column.}
\label{tab:data}
\end{table}

\subsection{Evaluation}\label{sec:evaluation}

We evaluate all models on five downstream protocols spanning retrieval, classification, similarity, and probing.

\textbf{Text-to-music retrieval.} We report mean reciprocal rank (MRR) on MusicCaps~\cite{agostinelli2023musiclm} (\num{2858} text-audio pairs from the official test set; free-text captions over AudioSet clips) and Song Describer~\cite{manco2023songdescriber} (\num{746} text-audio pairs from the official validation set; crowd-sourced expert descriptions).
Because MusicCaps draws from AudioSet, external baselines trained on AudioSet-derived audio are not strictly held-out on this benchmark.

\textbf{Zero-shot classification.} We compute accuracy on GTZAN~\cite{tzanetakis2002gtzan} (\num{10} genres, \num{1000} clips) and FMA-Small~\cite{defferrard2017fma} (\num{8} genres, \num{8000} clips) using cosine similarity between audio embeddings and text-encoded genre labels, prompted with \textit{``This is a [genre] song.''}

\textbf{Audio similarity.} DimSim~\cite{Lee2019MusicSimilarity} provides human-annotated pairwise similarity judgments. We report accuracy of the model's similarity ranking against the ground-truth ordering.

\textbf{MLP probing.} We freeze the audio encoder and train an MLP classifier on MagnaTagATune (MTT)~\cite{law2009mtt} and the three MTG-Jamendo subsets (Genre, Instrument, and Mood)~\cite{bogdanov2019mtgjamendo} for multilabel classification (macro mAP), and MGPHot for regression (macro RMSE)~\cite{ramoneda2026benchmarking}.
The probe is a two-layer MLP with a \num{512}-dimensional hidden layer, ReLU activation, and dropout (\num{0.2}) applied before each linear layer.
We optimize with AdamW at learning rate \num{1e-4}, batch size \num{64}, and cosine annealing with \num{2}\,k warm-up steps over \num{20}\,k training steps, and select the checkpoint with the lowest validation loss for testing.

\section{Experiments}\label{sec:experiments}

We start from the training setup of Section~\ref{sec:method} and conduct experiments varying one axis at a time.

\subsection{Training data analysis}\label{sec:exp_data}

We compare the two review-derived caption styles (quotes vs.\ structured) and their interaction with existing music and sound corpora.

\begin{table}
\centering
\footnotesize
\setlength{\tabcolsep}{4pt}
\begin{tabular}{rp{7cm}}
\toprule
\multicolumn{2}{c}{\textit{baseline+}\ourQuotes\ vs.\ \textit{baseline}} \\
\cmidrule(lr){1-2}
$\Delta$\,rank & Query \\
\midrule
\multicolumn{2}{l}{\textit{MusicCaps}} \\
$+$1951 & This music is instrumental. The tempo is slow with the musician plucking a single string of a ukelele. This audio is of a Ukelele being tuned. \\
$+$1869 & This audio recording features a crickets sound effect, sea waves sound effect and mellow synth pad chords in the background, followed by shimmering tambourine and muffled snare\ldots \\
$+$1822 & This piece is a live performance of dancers playing the tambourine over a rock music piece in the background. The background piece has a female vocal and an electric guitar\ldots \\
\midrule
\multicolumn{2}{l}{\textit{Song Describer}} \\
$+$291 & A rock song with a slow base well marked by drums and distorted guitars. \\
$+$282 & A power-pop song with a lot of idiosyncracies like flutes, a kid's choir, and guitar solo played backwards. \\
$+$246 & Male vocalist with a raspy voice singing over melancholic piano chords and drums increasing in intensity, with a slighty dissonant chorus featuring distorted guitars. \\
\bottomrule
\end{tabular}
\caption{Top-3 queries with the largest rank improvement when adding review supervision. $\Delta$\,rank = rank under review-augmented model $-$ rank under baseline. Larger is better. Captions are truncated to fit.}
\label{tab:top_review_gains}
\end{table}

\textbf{Combining with existing corpora.} 
We isolate each review corpus on top of a \emph{baseline} combining LPMC, M4-RAG, FS, and PSE.
Table~\ref{tab:data} shows that \ourQuotes\ or \ourStruct\ captions alone fall below the baseline on every metric: reviews alone do not match the breadth of the four-corpus mix.
Adding them on top, however, raises Song Describer MRR by \num{+2.8} (\textit{baseline+}\ourStruct) and \num{+3.7} (\textit{baseline+}\ourQuotes), while MusicCaps and zero-shot move only marginally.
Table~\ref{tab:top_review_gains} lists the largest per-query rank gains: review supervision unlocks queries that mix narrative framing, ambience, on-stage description, and nuanced instrumentation, all categories tag- and short-caption-style supervision leaves uncovered.
Adding \ourQuotes\ also consistently outperforms adding \ourStruct, which we attribute to its more narrative register, though prompting strategy and LLM vary (Section~\ref{sec:data}), so the two factors cannot be fully separated.
We use it as the reference corpus for the rest of the paper.


\begin{figure}
\centering
\includegraphics[width=\linewidth]{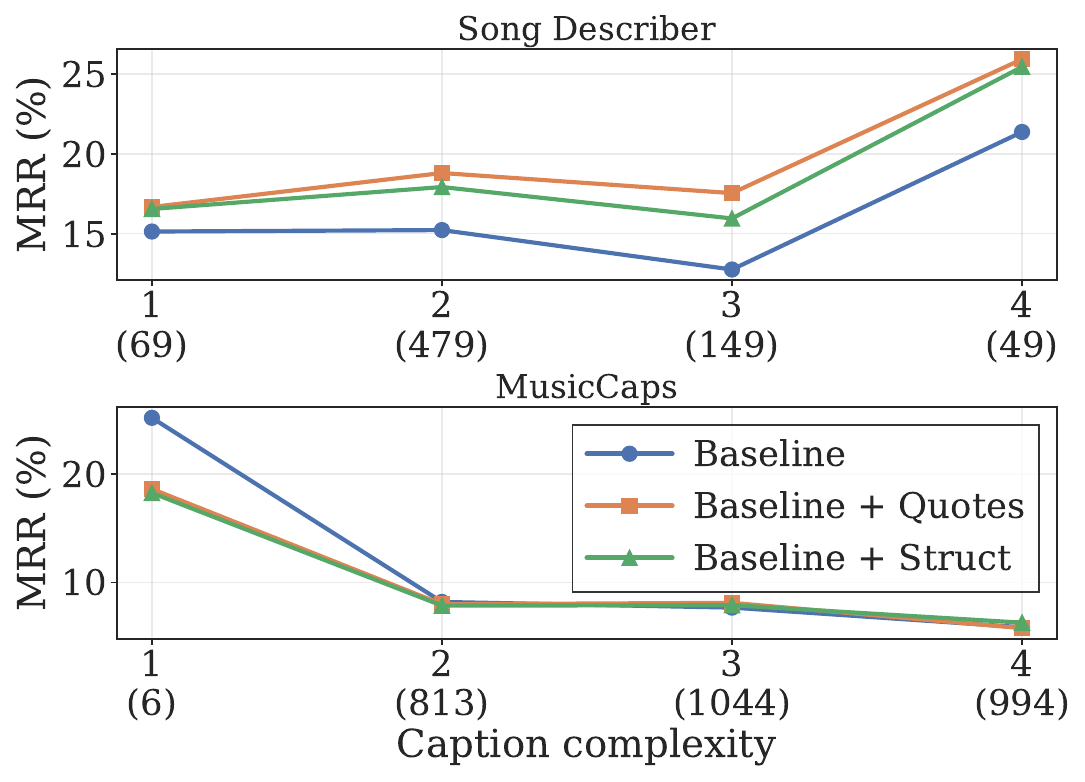}
\caption{Retrieval MRR (\%) by caption complexity. Complexity is a four-point LLM-judge score measuring how much descriptive content beyond bare tags each caption carries.
Parentheses indicate slice support.
}
\label{fig:complexity}
\end{figure}

\textbf{Lexical-slice analysis.}
To test whether the aggregate retrieval gains mask systematic variation across query types, we stratify each evaluation query by \emph{caption complexity} into four levels: 1) purely tag-expressible, 2) approximable by tags, 3) some narrative, and 4) largely narrative, evaluative, figurative, or context-dependent.
The partition is produced by Llama3-70b prompted with a rubric defining each level.
Figure~\ref{fig:complexity} plots MRR against complexity for models trained on the \emph{baseline} and the two review-augmented corpora.
The two datasets respond very differently: review supervision lifts Song Describer MRR at every complexity level, with the largest gains on score~3 and~4 captions, while MusicCaps MRR remains essentially flat, consistent with the aggregate gap in Table~\ref{tab:data}.

The top-gain queries in Table~\ref{tab:top_review_gains} combine narrative cues (``starts out with\ldots second half\ldots kicks in''), evaluative adjectives (``raspy'', ``melancholic'', ``dissonant''), and scene framing, suggesting the asymmetry is driven by the kind of non-tag content each dataset carries.
We make this precise with a register diagnostic over score~$\geq$~3 captions, using two trigger lists: a \emph{review-style} list (narrative, evaluative, scene framing: ``starts with\ldots'', ``soundtrack to\ldots'', ``feels like\ldots'') and a \emph{recording-style} list (``low quality'', ``amateur recording'', ``mono'', ``reverb'').
On Song Describer score~$\geq$~3, \SI{46.5}{\percent} of captions hit a review-style term and only \SI{3.0}{\percent} hit a recording-style term.
MusicCaps inverts this with \SI{46.8}{\percent} recording-style and \SI{28.5}{\percent} review-style, reflecting MusicCaps' musicologist-led annotation protocol.
We conclude that the benefits of our data concentrate on complex queries with narrative cues, evaluative adjectives, and scene framing, the register dominating Song Describer's captions and largely absent from MusicCaps.\footnote{The code and rubrics for the lexical-slice analysis and register diagnostic are included in the project repository.}




\begin{table}
\centering
\footnotesize
\setlength{\tabcolsep}{3pt}
\begin{tabular}{lccccc}
\toprule
 & \multicolumn{2}{c}{\textbf{Retrieval}} & \multicolumn{2}{c}{\textbf{ZS Class.}} & \textbf{Sim.} \\
\cmidrule(lr){2-3} \cmidrule(lr){4-5} \cmidrule(lr){6-6}
\textbf{Layer} & MuCaps & SongD. & GTZAN & FMA-S & DimSim \\
 & MRR$\uparrow$ & MRR$\uparrow$ & Acc.$\uparrow$ & Acc.$\uparrow$ & Acc.$\uparrow$ \\
\midrule
Layer 12 & 7.3 & 18.8 & \textbf{87.1} & \textbf{55.5} & 74.3 \\
Layer 6 & 6.5 & 17.9 & 83.5 & 55.0 & 82.0 \\
All layers & \textbf{7.8} & \textbf{19.3} & 85.4 & \textbf{55.5} & \textbf{83.0} \\
\midrule
 & \multicolumn{5}{c}{\textbf{MLP Probing}}  \\
\cmidrule(lr){2-6}
\textbf{Layer} & MTT & J.Genre & J.Instr. & J.Mood & MGPHot \\
 & MAP$\uparrow$ & MAP$\uparrow$ & MAP$\uparrow$ & MAP$\uparrow$ & RMSE$\downarrow$ \\
\midrule
Layer 12 & 43.4 & \textbf{21.7} & \textbf{17.1} & \textbf{15.2} & 0.162 \\
Layer 6 & \textbf{44.7} & 20.5 & 15.0 & 14.3 & 0.162 \\
All layers & 44.5 & 21.6 & 16.7 & \textbf{15.2} & \textbf{0.161} \\
\bottomrule
\end{tabular}
\caption{Audio encoder layer selection. All models use frozen text encoder and InfoNCE loss on \textit{baseline+}\ourQuotes\ data. 
\textbf{Bold} = best per column.}
\label{tab:layers}
\end{table}

\subsection{Layer selection}\label{sec:exp_layer}

The standard practice on CLAP models is to project the last layer of the audio encoder into the shared space.
Inspired by the frequent observation that intermediate layers sometimes produce more informative features~\cite{castellon2021codified, alonso2025omarrq}, we compare three configurations for the \num{12} transformer layers of the OMAR-RQ encoder: final-layer only, mid-layer (layer~6), and a learned combination of all layers.
Table~\ref{tab:layers} shows that switching from the final layer to a learned weighted sum of all OMAR-RQ layers improves MusicCaps and Song Describer MRR by \num{0.5} absolute and DimSim accuracy by \SI{8.7}{pp} over the layer-12 baseline, while probing metrics remain comparable.

\begin{figure*}
    \centering
    \includegraphics[width=1\linewidth]{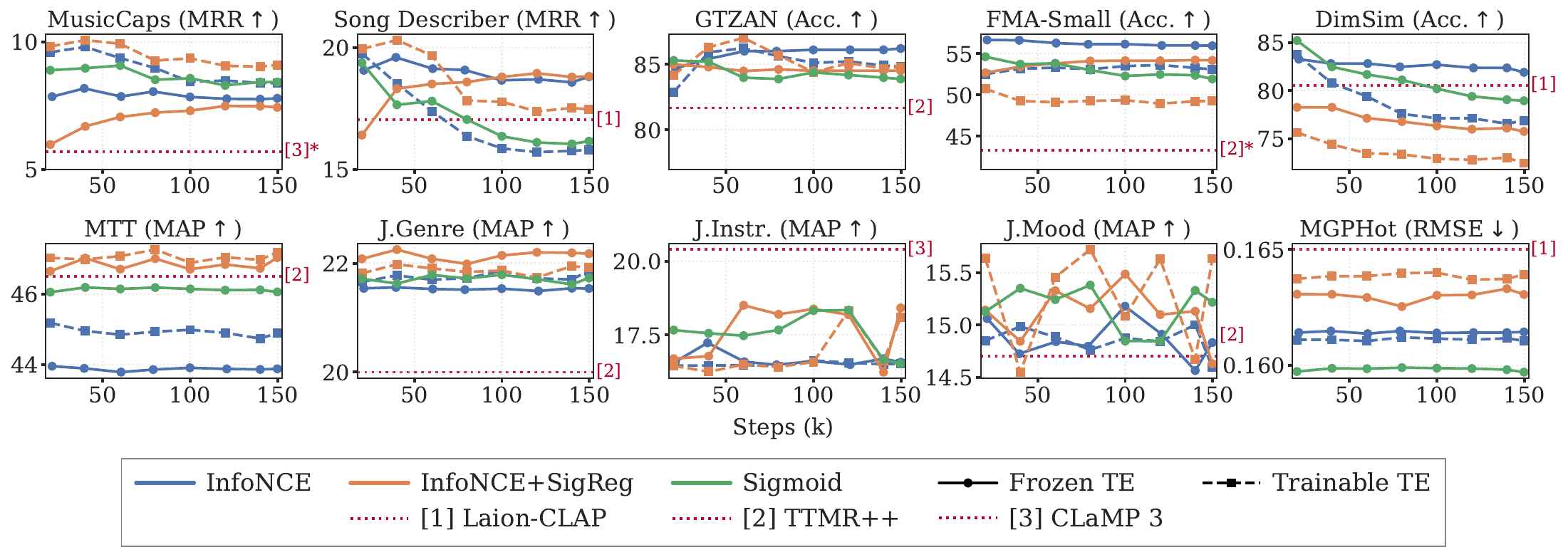}
    \caption{
Downstream performance over model training steps. Dotted crimson lines mark the strongest external baseline per task, evaluated using our pipeline. 
All our models use all OMAR-RQ layers and \textit{baseline+}\ourQuotes\ data.
\emph{LeJEPA} is omitted: it tracks \emph{InfoNCE+SigReg} on probing but underperforms elsewhere.
$^{*}$On MusicCaps and FMA-Small, we exclude Laion-CLAP and TTMR++, and Laion-CLAP respectively due to train-test overlap.
    }
    \label{fig:dynamics}
\end{figure*}

\subsection{Training objective}\label{sec:exp_objective}

Fixing the layer and data configuration (all layers, \textit{baseline+}\ourQuotes), we compare the symmetric InfoNCE loss (Equation~\ref{eq:infonce}) with three other objectives.

\textbf{Sigmoid loss}~\cite{zhai2023sigmoid} replaces softmax with per-pair binary cross-entropy.
With $\sigma(\cdot)$ the logistic sigmoid, $y_{ij} = 2\,\mathbb{1}[i{=}j]-1$, $t$ a learnable temperature, $b$ a learnable bias, and $\ell^{xy}_{ij} = \log\sigma(y_{ij}(t\,\mathbf{x}_i^\top\mathbf{y}_j + b))$:
\begin{equation}
  \mathcal{L}_{\text{Sigmoid}} =
    -\frac{1}{2N}\sum_{i,j}\!\left(\ell^{at}_{ij} + \ell^{ta}_{ij}\right).
\end{equation}

\textbf{LeJEPA}~\cite{balestriero2025lejepa}
is a negative-free self-supervised method combining an invariance loss with a geometrical regularization term.
Letting $\mathbf{z}_{i,m} \in \mathbb{R}^{d}$ denote the embedding of sample $i$ for modality $m \in \{a, t\}$, SigReg
  encourages $\mathbf{z}_{i,m}$ to be approximately Gaussian by minimizing a normality test $\mathcal{T}$ over $K$ random
  unit-vector projections $\mathbf{v}_k \sim \mathcal{U}(\mathcal{S}^{2d-1})$:

\begin{equation}
  \mathcal{L}_{\text{SigReg}} =
    \frac{1}{M}\sum_{m=1}^{M}\frac{1}{K}\sum_{k=1}^{K}
      \mathcal{T}\!\left(\bigl\{\mathbf{v}_k^\top \mathbf{z}_{i,m}\bigr\}_{i=1}^{N}\right),
\end{equation}
where we use the Epps-Pulley test as $\mathcal{T}$, $M = 2$ and $K = 17$.
For LeJEPA, our invariance term is the cosine distance between modality embeddings $\mathbf{z}_{i,m}$ and the $\ell_2$-normalized average across modalities $\bar{\mathbf{z}}_i$, with $\lambda = 0.05$:
\begin{equation}
  \mathcal{L}_{\text{LeJEPA}} = \frac{1 - \lambda}{M N}\sum_{m,i}\!\left(1 - {\mathbf{z}}_{i,m}^{\top} \bar{\mathbf{z}}_i\right) + \lambda\,\mathcal{L}_{\text{SigReg}}.
\end{equation}

\textbf{InfoNCE + SigReg.}
Finally, we combine the standard contrastive objective with SigReg ($\lambda = 0.05$):
\begin{equation}
  \mathcal{L}_{\text{InfoNCE+SigReg}} =
    (1 - \lambda)\mathcal{L}_{\text{InfoNCE}} + \lambda\,\mathcal{L}_{\text{SigReg}}.
\end{equation}

We use the largest batch size fitting in GPU memory: 3072 with a frozen text encoder (\emph{InfoNCE}, \emph{InfoNCE+SigReg}, \emph{Sigmoid}) and 1024 with a trainable one (\emph{InfoNCE}, \emph{LeJEPA}, \emph{InfoNCE+SigReg}).
Figure~\ref{fig:dynamics} shows the downstream dynamics over training steps, separating stable trajectories from noisy ones whose differences fall within run variance.
Most models peak early (\num{20}--\num{60}\,k steps) even as train/val losses keep decreasing through \num{150}\,k, and then degrade on non-probing tasks, most visibly on models with the \emph{Sigmoid loss} or a trainable text encoder.
Training the text encoder helps text-to-music retrieval but hurts similarity (DimSim) and gives mixed zero-shot results (worse on FMA-Small, while \emph{InfoNCE+SigReg} tops GTZAN).
Probing shows no monotonic degradation: MTT, MTG-Jamendo Genre, and MGPHot are stable, while MTG-Jamendo Instrument and Mood oscillate within a \num{2} MAP-point span.
\emph{LeJEPA} and \emph{InfoNCE+SigReg} consistently beat \emph{InfoNCE} on all classification tasks, aligning with the reported benefits of isotropic Gaussian representations for probing~\cite{balestriero2025lejepa}, though \emph{LeJEPA}'s lack of negatives penalizes every non-probing task.

\subsection{Comparison with external baselines}\label{sec:comparison}

We compare our models against three open CLAP-style external baselines, evaluated with our downstream pipeline.
\textbf{Laion-CLAP}~\cite{wu2023largescale} is a general text-audio contrastive model trained on \num{2.5}\,M (audio, text) pairs pooled from AudioSet, Freesound, and other web sources. We use the checkpoint \textit{music\_speech\_epoch\_15\_esc\_89.25}.
\textbf{TTMR++}~\cite{doh2024enriching} is a music-focused CLAP trained with LLM-generated captions derived from tags over the LPMC, AudioSet (Music), Music4All~\cite{santana2020music4all}, FMA, and MusicCaps datasets.
\textbf{CLaMP3$_{saas}$}~\cite{wu2025clamp} is a cross-modal music encoder in its ``symbolic-as-audio'' setting, trained on the full M4-RAG.
We did not consider other open-weights models such as MuQ-MuLan~\cite{muqmulan2024}, for which training data is undisclosed.
On MusicCaps, GTZAN, FMA-Small, MTG-Jamendo Genre, MTG-Jamendo Mood, and MGPHot, our models perform better than the strongest external baseline on almost every configuration and training step.
In contrast, none of our models surpasses the strongest external baseline on MTG-Jamendo Instrument, and only \emph{InfoNCE+SigReg} surpasses it on MTT, whose top-50 taxonomy is rich in instrument tags.
Consistent with Section~\ref{sec:exp_data}, album-level reviews cover broad aspects (genre, scene, mood) better than production-style ones (instrumentation).
On Song Describer and DimSim, our models perform closer to the external baselines and are only superior for certain objectives and training-step settings.





\section{Conclusions}\label{sec:conclusions}

We present \ourDataset, a dataset of \num{245346} captions distilled from AllMusic album reviews and supplementary metadata.
Combining it with existing music and general-sound datasets, learning a weighted sum over all audio-encoder layers, and adding a SigReg regularization term, our CLAP model leads on text-to-music retrieval and MLP and zero-shot classification, except for MTG-Jamendo Instrument.
To support open research, we release \ourDataset, model weights, and code for non-commercial research purposes.
Future work includes deepening the study of geometrical regularization in CLAP models, and applying \ourDataset{} to music captioning.

\section{Acknowledgements}\label{sec:acknowledgements}

This work is supported by the ``Cátedra IA y Música'' project (TSI-100929-2023-1), funded by the Secretaría de Estado de Digitalización e Inteligencia Artificial, the European Union-Next Generation EU funds and BMAT Music Innovators. And by the ``IMPA'' project (PID2023-152250OB-I00) funded by MCIU/AEI/10.13039/501100011033/FEDER, UE.
We thankfully acknowledge the computer resources at MareNostrum and the technical support provided by Barcelona Supercomputing Center (Activities \mbox{IM-2025-1-0026} and \mbox{IM-2025-3-0043}).

\section{AI usage statement}\label{sec:ai_usage}

LLMs are a core component of this work: \mbox{Qwen~2.5-32B} generates the \ourQuotes\ captions, while \mbox{Llama~3-70B} generates the \ourStruct\ captions and the caption-complexity annotations (Sections~\ref{sec:data} and~\ref{sec:exp_data}).
Beyond this methodological use, we employed autocomplete and agentic coding tools for the experiment implementation, and LLMs to review and improve the writing.

\bibliography{main}

\end{document}